\documentclass[pre, %
floatfix,
12pt,
superscriptaddress,
amsmath,amssymb,
aps, 
preprint]{revtex4-2}

\usepackage{graphicx}
\usepackage{dcolumn}
\usepackage{bm}

\usepackage[inkscapelatex=false]{svg}

\usepackage{comment}

\begin{document}
	
	
	\title{Scaling of poroelastic coarsening and elastic arrest in crosslinked gels}%
	\author{Samuel A. Safran}
	\affiliation{Department of Chemical and Biological Physics, Weizmann Institute of Science, Rehovot 76100, Israel}
    \email{sam.safran@weizmann.ac.il}
\date{\today}

\begin{abstract}
Recent experiments on crosslinked gels quenched from solvent-rich to solvent-poor conditions show solvent-rich domains embedded in a gel-rich matrix. These domains coarsen and then undergo kinetic arrest at micron scales for hours, before macroscopic drainage to equilibrium over even longer times. Motivated by these observations, we develop a minimal model that couples capillarity-driven Darcy permeation to the viscoelastic-to-elastic crossover of the polymer network. In the viscoelastic regime, the Young--Laplace traction at curved solvent--gel interfaces generates a pressure gradient in the solvent pores of the gel that drives solvent flow and coarsening. In the elastic regime, the same interfacial traction is balanced by elastic stress. This force balance eliminates pressure gradients in the solvent-filled pores of the gel, removing the Darcy driving force and arresting coarsening. Using the kinetic criterion $t(\lambda_{\rm arrest}) \sim \tau_{\rm el}$, we predict stiffness-dependent coarsening and arrest laws. For melt-like, polymer-rich gels, $\lambda(t)\sim G^{-1/2} t^{1/4}$ and $\lambda_{\rm arrest}\sim G^{-1/2}$. For low polymer fractions where the mesh size controls transport, $\lambda(t)\sim G^{-1/3} t^{1/3}$ and $\lambda_{\rm arrest}\sim G^{-1/3}$. The predicted $G^{-1/2}$ arrest scaling for melt-like gels agrees with experiment.
\end{abstract}

\maketitle

	\section{\label{sec:introduction}Introduction}
Recent experiments \cite{FernandezRico2024NatMat} on phase separation of PDMS (polydimethylsiloxane)  cross-linked polymers swollen by an organic solvent (a polymer gel system) have shown results that are strikingly different from our simple understanding of phase separation in simpler, two-component systems.   The cross-linked polymer PDMS gel is initially prepared at a higher temperature with a certain fraction of solvent in equilibrium. The temperature is then lowered, and the equilibrium solvent fraction within the gel decreases. 

At its lower temperature state, the gel initially contains excess solvent due to its initial high-temperature preparation. After an initial period of spinodal decomposition and nucleation, mesoscale domains of solvent-rich regions form in the gel, surrounded by gel that is polymer-rich and has less solvent than its initial, high temperature state. The domain structure coarsens (with small domains shrinking and large ones growing), and this process is reminiscent of Lifshitz-Slyozov-Wagner (LSW) \cite{LifshitzSlyozov1961, Wagner1961,Bray1994},  coarsening in the phase separation of two-component systems.
Strikingly, in these experiments the coarsening undergoes long-lived kinetic arrest at submicron or micron length scales, depending on the rigidity of the cross-linked gel. This mesoscale structure remains stable for several hours. However, over much longer times,  the excess solvent slowly redistributes and migrates to the external solvent reservoir. The gel then relaxes to a homogeneous low-temperature state with a lower solvent fraction than in its initially prepared state.

The origin of this metastable, elastic  arrest has interested several scientific groups. The main experimental observation \cite{FernandezRico2024NatMat} is that the long-time mesoscale domain size depends on $1/\sqrt{G}$, where $G$ is the shear modulus of the gel, as measured in six gels with $10\text{kPa}\lesssim G \lesssim 400\text{kPa}$.
The coarsening kinetics were treated numerically using a Cahn-Hilliard type model (see Supplementary Information (SI) of \cite{FernandezRico2024NatMat} ), but no experimental quantification was presented. However, the paper states that for the more rigid gels, no coarsening was observed on the \textit{time scale of the measurements}, while for the three softest gels, coarsening of the domains was observed.

A paper with theory and model calculations of phase separation and coarsening in a viscoelastic gel by Curk and Luijten \cite{CurkLuijten2023PNAS} extended LSW-type Ostwald ripening driven by capillarity (interfacial tension $\gamma$) by incorporating elastic stresses and their viscoelastic relaxation in the surrounding matrix (see Appendix \ref{app:curk}). Their approach is based on diffusion-controlled evaporation–condensation through a viscoelastic solid, modified by elastic and viscoelastic effects.  They do not include poroelastic (Darcy) solvent flow through a deformable, porous network, which provides the dominant dissipative mechanism for relaxing capillary pressure in our approach and leads to the elastic arrest prediction of scaling with $1/G^{1/2}$, in agreement with the experiments \cite{FernandezRico2024NatMat} (see Appendix~\ref{app:lambdastar}).  A previous, numerical calculation by Onuki and Puri \cite{OnukiPuri1999}  focused on  a two-dimensional crosslinked gel with (implicit) solvent-gel friction. In their numerical solutions the onset of elastic arrest occurred at a time that depended on the value of the modulus (see Appendix for further discussion).  A more recent theory of phase separation in elastic networks \cite{PaulinSoftMatter2026}, based on coupled chemical-potential gradients and elastic stress, concluded that local elasticity alone cannot arrest coarsening, and that non-local elastic interactions are required (see Appendix \ref{app:onukipuri}). By contrast, our approach focuses explicitly on solvent \textit{flow} through the gel coupled to the polymers' poroelastic response. By treating solvent permeation in the gel (Darcy flow) and the elastic back-reaction directly—rather than incorporating elasticity solely through constitutive stress–energy relations—we demonstrate that elastic arrest of coarsening can occur even for linearly elastic gels, as explained below.
 
Other theories \cite{QiangLuoZwicker2024PRX,MannattilDiamantAndelman2025PRL,FernandezRico2026SoftMatter,Wei2020PRL} proposed to explain these phenomena using equilibrium models, unrelated to coarsening and poroelastic flow.  However, these studies do not account for:
(a) Coarsening dynamics and its relation to solvent flow in the porous and deformable gel. 
(b) The fact that the system ultimately reaches equilibrium as a homogeneous gel which suggests slow drainage of solvent into the reservoir.  In order for the solvent-rich domain scale to be magnified from the nm mesh-size of the gels of \cite{FernandezRico2024NatMat} to the observed micron scale, these theories require a magnification factor due to inhomogeneities \cite{QiangLuoZwicker2024PRX,MannattilDiamantAndelman2025PRL} or an anomalously large stretchability of the gel \cite{Wei2020PRL}. In addition,  model of \cite{Wei2020PRL}, the predicts that the domain varies as $1/G^{1/3}$ instead of the square root behavior of the experiments \cite{FernandezRico2024NatMat}.  More importantly,  micron-scale solvent domain sizes are related in \cite{Wei2020PRL} to the assumption that that the solvent-rich domain is still a coherent, but ultra-dilute gel; the stretchability of the gel is something that should be tested experimentally.  These equilibrium models, as well as the dynamical models of \cite{OnukiPuri1999, CurkLuijten2023PNAS, PaulinSoftMatter2026} do not relate to the Darcy-like solvent flow in the gel, driven by the capillary pressure of the solvent-rich domain. As we explain below, in our theory of coarsening due to Darcy flow and its arrest in the elastic state of the gel, the magnification factor arises from the large viscosity ratios of the polymers and the solvent.

Here, we present a unified scaling theory of both domain  coarsening and elastic arrest in crosslinked gels which exhibit crosslink-controlled \textit{elastic} behavior at long times and a  \textit{viscoelastic} response activated by solvent flow through the network at intermediate times. Solvent flow (through ``Darcy's law'') is driven by the pressure gradients within the solvent pores of the gel \cite{TanakaFillmore1979,Biot1956,HuSuo2012PoroviscoelasticElastomericGels}.  It occurs at time scales for which poroelastic dissipation, acting within the viscoelastic regime of the polymer network,  controls the rate of coarsening.  In contrast, the onset of long-time \textit{elastic} response governs kinetic arrest \cite{DoiOnuki1992,OnukiPuri1999,TanakaFillmore1979}, since as discussed below, the elastic equilibrium is that of a \textit{shear} stress with no pressure gradient in the gel and hence no solvent flow or coarsening.  When the characteristic coarsening time becomes comparable to the relaxation time separating the flowing, viscoelastic regime from the solid-like elastic regime of the crosslinked polymers, further capillary-driven domain evolution is arrested by the development of long-range elastic stress in the gel. At the level of the Laplace force balance at the solvent-rich domain boundary, the shear stress balances the capillary driving force; there is no pressure gradient in the solvent pores of the gel and no driving force for further solvent flow.
 The resulting arrested domain size is therefore  not set directly by microscopic gel length scales, but by a kinetic criterion and is amplified by a \emph{magnification factor} arising from the large separation between the poroelastic time governing solvent transport and the much longer time required for the onset of elastic response. During this extended viscoelastic window, solvent can be transferred over progressively larger distances, converting nanoscale gel properties into mesoscale domain sizes at arrest.

The complexity of the problem is simplified by our  scaling approach focused on the dimensionally correct dependence of the coarsening and elastic arrest in terms of the gel properties, without the fine distinctions of numerical factors or dimensionless factors of order unity \cite{deGennes1979}.  Thus, instead of equalities, we use the convention $\sim$ to mean ``scales as''.  The theory predicts that the coarsening length scales as $t^{1/4}$ for slow, poroelastic diffusion and that the arrest size of the domains scales as $1/\sqrt{G}$; the latter agrees with the experiments.  Finally, we note here that while the explanations below consider the case of shrinking, solvent-rich domains, our model and its scaling predictions apply to both shrinking and growing domains, which are of course related by conservation of solvent. When we discuss solvent flowing out of shrinking solvent-rich domains, the same physics applies to solvent being drawn into growing domains.

\section{\label{sec:model1}Methods}
\subsection{LSW coarsening} 
For contrast with the coarsening of solvent-rich domains in gels, we briefly review the scaling arguments that lead to domains that grow or shrink as $t^{1/3}$\cite{Bray1994}.  We consider a binary (AB) liquid that has been quenched from its mixed phase to a temperature where the equilibrium state is one of macroscopic phase separation. In general, each of those phases contains both A and B and for simplicity, we denote the phase by its majority component (e.g., B-rich is termed B). At early times, the system undergoes spinodal decomposition followed by nucleation of  domains of B, for example, of one liquid in the background of the other \cite{Bray1994, Debenedetti1996}. 

These domains coarsen \cite{Bray1994} and the kinetics of coarsening are based upon the conservation relation that equates the number of molecules $N_B$ leaving (for example) a shrinking B domain (and eventually being drawn into a growing  B domain), to their flux $J_{BA}$ at the  interface of the B domain with the background of A: $dN_B/dt = - J_{BA} S$, where the flux is evaluated locally at the domain boundary and is controlled by the driving force acting at that boundary.   For scaling purposes, we can ignore the vectors, keeping in mind that the flux and velocity of the B-molecules exiting the B-rich domain are in the direction normal to the domain boundary.  Here $S$ is the area of the shrinking domain which, for simplicity, we have taken to be spherical and henceforth we denote the time derivatives by a dot.  $\dot N_B$ is related to the composition difference $\Delta c$ between the two phases and the change in the domain interface position and the domain area.  The flux is the number of B molecules that exit  the B-rich domain, per unit area per unit time and is written in terms of the B-molecule velocity as: $J_{BA}=c_B \  v$, where  $c_B$ is the concentration (number density) of B molecules in the B-rich domain.  For a shrinking domain, $\dot N_B <0$ implies that $J_{BA}>0$; the velocity is directed away from the domain.   In scaling, we write: $\dot N_B \sim - \Delta c \, S \, \dot \lambda$ where  $\lambda$ is the size of the $B$ domain.  For a more detailed discussion of this and the following see Appendix \ref{app:unified_diffusion}.

The velocity  is driven by the chemical potential gradient of the molecules that exit the spherical domain; the value of this gradient evaluated at the domain surface  determines the interfacial flux.  The energy of this driving force is dissipated by the diffusion of the B molecules through the A-rich regions, with mutual diffusion constant $D$.  Thus,$J_{BA} \sim D c_B \nabla \mu$ . The chemical potential of a molecule in a spherical domain of size $\lambda$ is higher than that of a molecule at a flat interface by the amount $\delta \mu \sim  \ \gamma/(c_B )\lambda$ where $\gamma$ is the interfacial tension.  The factor of $c_B$ converts $\delta \mu/\lambda$ to an energy and we take all energies to be in units of $k_BT$.  

Using the conservation equation, we now write a scaling relation: 
$\Delta c \ \dot \lambda  \sim  - D \   \gamma/(\lambda \ \ell) $ where $\Delta c$ is the concentration difference of the B molecules in the B-rich phase and the A-rich phase. $\ell$ now is the extent of the chemical potential gradient in the direction  normal to the surface of the B domain, evaluated at the domain interface. LSW scaling for the domain growth is obtained by approximating $\ell$ by $\lambda$ which originates in the solution of the diffusion equation for the B molecules in the background of A, at the domain surface \cite{Bray1994} and Appendix \ref{app:unified_diffusion}.  This equation in the \textit{fast-diffusion} limit (applicable to molecular diffusion in a liquid relative to the coarsening times), results in an equation for the spatial profile of the B molecule concentration in the A-rich domain where $\lambda$ determines the scale of the gradient. Below, we generalize the approximation for the gradient to include diffusion. 

The result for a shrinking domain gives $\dot \lambda \sim - (D  /\Delta c) \, \gamma/\lambda^2$.  For scaling, we can take $\Delta c \sim c_B$ (not valid near a critical point) which gives the LSW scaling, $\lambda \sim \left( \ (D \ \gamma/c_B ) \ t \right)^{1/3}$, where we do not distinguish shrinking from growing vis a vis the scaling; the two are simply related by conservation of the B molecules. As discussed below, the equations for the chemical potential gradients can be transformed into equations for the osmotic pressure gradients of the B molecules in the A-rich phase.

\subsection{Generalizing LSW to coarsening of solvent-rich domains in  viscoelastic gels:} 
In the following, the gel properties such as interfacial tension with a solvent-rich domain, elastic modulus $G$, mesh size $\xi$   those that characterize the equilibrium gel.
We again begin with the conservation relation, $\dot N_B \sim - \Delta c_s \, S \, \dot \lambda = -J S$  where the flux is evaluated locally at the domain boundary and is controlled by the driving force acting at that boundary. $\Delta c_s$ is the difference between the solvent concentration in the solvent-rich domain and the gel.   The flux is given by  $\vec J=c_s \vec v$ where $c_s$ is the solvent concentration in the solvent-rich domain which, due to capillary pressure, releases solvent molecules into the gel with a velocity $\vec v$.  For a more detailed derivation of this and the following, see Appendix \ref{app:unified_diffusion}.

However, instead of diffusing in the polymer-rich phase, the solvent flows in the deformable, viscoelastic gel considered as a porous medium.  The flow velocity $\vec v$ is given by the Darcy relation \cite{deGennes1979, Biot1941, Biot1956,SkotheimMahadevan2004} that relates this velocity to the pressure gradient in the solvent pores of the gel: $\vec v = (k_0/\eta_D) \nabla p$.  Here $k_0$ is the permeability of the gel and $\eta_D$ is the viscosity appropriate to Darcy flow to the solvent  impeded by its friction with the polymer network. Because this transport is controlled by the hydrodynamic permeability of the solvent rather than
by polymer rearrangements, the network topology is effectively fixed on the Darcy-flow timescale: polymer configurations do not relax rapidly enough to eliminate the stress induced by solvent flow.   Physically, this reflects a separation of timescales between the relatively fast permeation of small solvent molecules through the gel and the much slower relaxation of stress carried by long-chain polymers.
In the Results, we discuss specific choices for these parameters. 

The pressure gradient entering Darcy’s law is the local gradient at the interface of the solvent-rich domain and the gel, whose magnitude is set by how the interfacial pressure perturbation spreads into the gel.  In incompressible \textit{fluid} systems  pressure gradients are proportional to  chemical potential gradients of the solvent molecules in the gel, by the thermodynamic relation,
 $\nabla \mu=\nabla p/c_B$; see Eq.~1.1.14 of Ref.~\cite{Onuki2002} and Appendix \ref{app:unified_diffusion}. The chemical potential at the solvent-rich domain surface is again given by capillarity, $\delta \mu = \gamma / (\lambda c_s)$ and the osmotic pressure jump at the interface (which acts as the mechanical pressure jump) is given by the Laplace law as $\delta p= c_s \delta \mu = \gamma/\lambda$. Finally, the conservation law and the Darcy relation used in the flux give the dynamical scaling relation for a shrinking domain: 
 \begin{equation}
 \dot \lambda \sim - \frac{k_0 \ c_s}{\eta_D} \nabla p \sim - \frac{k_0 \ \gamma }{\eta_D  \ \lambda \ \ell_p } 
 \label{eq:cons1}
 \end{equation}
where $\ell_p$ is the scale of the pressure gradient, evaluated at the interface of the solvent-rich domain and the gel, and where we have set $c_s/\Delta c_s$ to be a dimensionless constant of order unity for scaling purposes.

The solvent pressure in the gel decreases from its maximal value at the boundary of the shrinking domain via stress diffusion, known as poroelasticity \cite{SkotheimMahadevan2004, ChesterAnand2010}.  That both the stress and pressure propagate diffusively can be understood from the low-Reynolds number force balance in a gel where pressure gradients are balanced by viscoelastic, polymer stress gradients. At relatively early times, when the poroelastic diffusion length \(\ell_p\) is smaller than the domain size \(\lambda\), the pressure gradient at the domain boundary can be estimated as \(\nabla p \sim \delta p / \ell_p\); this approximation captures the transient buildup of the interfacial gradient before it reaches its long-time, saturated value.  As defined by poroelastic diffusion \cite{TanakaFillmore1979}:
\begin{equation}
	\ell_p(t)=\sqrt{D_p\, t},
\end{equation}
where \(D_p \sim k_0\, G / \eta_{ps}\) is the poroelastic diffusion constant, with \(k_0\) the permeability of the gel, \(G\) its elastic modulus, and \(\eta_{ps}\) the viscosity associated with the  viscoelastic response (the subscript ps referring to the poroelastic regime) of the coupled polymer--solvent system activated by solvent flow at intermediate times.  
This description applies only until \(\ell_p\) reaches the domain size \(\lambda(t)\).   As shown in Appendix \ref{app:unified_diffusion}, this is because the finite domain size, together with the Laplace pressure boundary condition at the domain interface, fix both the magnitude and the spatial extent of the pressure gradients.  Those are substantial only for distances from the domain smaller or equal to $\lambda$. Beyond this point, the solution of the poroelastic diffusion equation has evolved so that the pressure gradient evaluated at the domain interface no longer changes with time.

We note that the elastic modulus enters the poroelastic diffusion constant even when the gel is in its \textit{viscoelastic} regime because the modulus
controls the \textit{initial} magnitude of the viscoelastic stress.  In the viscoelastic regime, this stress does not persist
indefinitely, but it governs the rate at which solvent redistributes while it is
present. By contrast, in the \textit{elastic} regime, the stress persists up to very long times as discussed below.

We note that most treatments of poroelasticity take the crosslinked gel to have
an elastic (as opposed to viscoelastic) stress response
\cite{TanakaFillmore1979,HuSuo2012}.  This indeed applies for times greater than
the intrinsic viscoelastic crossover time of the gel, $\tau_{\mathrm{el}}$,
which separates the early-time \textit{viscoelastic} stress response,  from the long-time \textit{elastic} response
\cite{Ferry,RubinsteinColby}.  For $t \gg \tau_{\mathrm{el}}$, the shorter-time
polymeric degrees of freedom have equilibrated, the crosslinked polymers behave
elastically, and dissipation associated with stress redistribution is controlled
primarily by solvent flow relative to a static polymer network.  In this regime,
the viscosity entering stress diffusion is well approximated by the Darcy
viscosity, so that $\eta_{ps} \approx \eta_D$.

During coarsening, however, the gel is responding dynamically to stresses imposed
by solvent being released from shrinking domains and accumulated in growing
domains.  These stresses act as localized perturbations to a gel that is
otherwise close to its equilibrium swollen state. Because these perturbations are time dependent and spatially localized, they do not probe the static, long-time elastic response of the network, but instead drive transient polymer stress relaxation. Permanent crosslinking suppresses macroscopic polymeric flow but does not eliminate internal polymer dissipation; even dry covalently crosslinked networks exhibit viscoelastic stress relaxation at intermediate times due to polymeric (individual chain and local, transient cooperative) relaxation modes , before reaching an elastic plateau \cite{StrangeFletcherTonsomboonBrawnZhaoOyen2013SeparatingPoroviscoelastic}. For this reason, we envision
the gel responding viscoelastically on the timescale of coarsening, before
elastic arrest intervenes and halts the dynamics at long times (on the scale of
hours).  The viscoelastic response of the polymer, which applies even to permanently crosslinked gels, arises from the time-dependent, local strains in the gel due, for example, from the solvent being injected by a shrinking domain. These strains generate time-dependent polymer stresses that excite internal viscoelastic relaxation modes associated with frictional rearrangements of polymer segments between crosslinks. 
The local and transient polymer rearrangements contribute to stress
relaxation, and the relevant dissipation expressed by the effective  stress-relaxation viscosity $\eta_{ps}$ reflects both solvent--polymer friction
and the intrinsic viscoelastic dynamics of the polymer-rich phase
\cite{RubinsteinColby,Onuki1997}. 

Accordingly, we take the viscosity entering poroelastic diffusion during
coarsening to be a poroelastic viscosity $\eta_{ps}$ that is distinct from the
Darcy viscosity $\eta_D$ which characterizes solvent flow in
response to pressure gradients in a static network.   In contrast, $\eta_{ps}$ governs
the coupled relaxation of pressure and polymer stress when the network itself is
dynamically responding. This distinction must be made even in a permanently crosslinked gel. There too, solvent transport generates strains on timescales comparable to polymer stress relaxation. In general, therefore, $\eta_{ps} \neq \eta_D$ for
$t \ll \tau_{\mathrm{el}}$.  We comment further on the implications  of this distinction in the
Results and Discussion.

When poroelastic diffusion governs the relaxation of pressure gradients,
Eq.~\ref{eq:cons1} can be written in the scaling form
\begin{equation}
\dot{\lambda} \sim - \frac{k_0 \gamma}{\eta_D\,\lambda\,\ell_p},
\label{eq:cons2}
\end{equation}
where $\ell_p$ denotes the characteristic length over which pressure relaxes in
the gel.  The implications of this form, and the behavior of $\ell_p$ in
different regimes, are discussed in the Results.

We consider two limiting regimes of poroelastic dissipation. 
\emph{(i) Fast poroelastic diffusion.}  If stress and pressure relax rapidly in
the viscoelastic gel, the pressure gradient induced by capillary driving decays
over a length scale of order the domain size.  In this limit, coarsening follows
kinetics analogous to the Lifshitz--Slyozov--Wagner case, with
$\lambda \sim t^{1/3}$.  
\emph{(ii) Slow poroelastic diffusion.}  If poroelastic relaxation is slow, the
pressure gradient decays over a poroelastic length that grows diffusively,
$\ell_p \sim \sqrt{D_p\,t}$.  Substituting this into Eq.~\ref{eq:cons2} yields the
scaling law
\begin{equation}
\lambda \sim
\left(\frac{k_0\,\gamma}{\eta_D\,D_p^{1/2}}\right)^{1/2} t^{1/4}.
\label{eq:onequarter}
\end{equation}
As above, shrinking and growing domains obey the same scaling by solvent
conservation.  In this slow-poroelastic regime, the coarsening law exhibits an
explicit dependence on the gel modulus through the poroelastic diffusion
constant, $D_p \sim G$, giving $\lambda \sim G^{-1/4} t^{1/4}$.  The extent to
which the prefactor $\sqrt{k_0 \gamma / \eta_D}$ introduces additional modulus
dependence is discussed further in the Results.

The use of the diffusion length $\ell_p=\sqrt{D_p \ t}$ for the slow poroelastic limit is only appropriate if $\ell_p<\lambda(t)$; relaxation of the gel on scales larger than the domain size $\lambda(t)$ is not relevant for solvent release n into the gel by a shrinking domain of size $
\lambda$ (see Appendix \ref{app:unified_diffusion}). Thus, the slow poroelastic limit crosses over to the fast limit at a crossover domain size.  Equating $\ell_p\sim\lambda(t)$ yields a time-independent crossover length
\begin{equation}
\lambda_\times\sim \,\frac{\gamma}{G}\,\frac{\eta_{ps}}{\eta_D},
\label{eq:xover}
\end{equation}
which magnifies the elastocapillary length $\gamma/G$ (which is of the order the mesh size of the gel) by a viscosity ratio; numerical estimates are presented in the Results section. The crossover length  can be large since the Darcy viscosity $\eta_D$ accounts for the solvent viscosity and friction with the instantaneous state of the polymers of the gel while the poroelastic viscosity in the \textit{viscoelastic} regime of the gel, where coarsening takes place, accounts for polymer-polymer friction and local chain relaxation  When the domain size $\lambda(t)<\lambda_\times$ the slow limit applies with coarsening scaling as $t^{1/4}$.  In the opposite limit, the pressure gradients decay as $1/\lambda(t)$ which is appropriate in the fast poroelastic limit where the domain size scaling is $t^{1/3}$.

\section{Elastic arrest}
\label{sec:elastic_arrest_rewritten}

So far, we have focused on the capillary driving force for coarsening. 
At early times, the coarsening gel is in the \textit{viscoelastic} regime, 
where polymers respond locally and transiently to the stresses generated by 
capillary tractions (normal stresses) at curved solvent--gel interfaces. 
In this regime, there is no spatially \textit{long-ranged} and temporally \emph{persistent} elastic response; 
transient polymer stresses therefore act primarily as a source of dissipation, 
slowing---but not halting---the capillarity-driven coarsening dynamics. 
Mechanically, the Young--Laplace condition prescribes a normal traction (normal stress) at the domain boundary; 
because a long-ranged, persistent deviatoric stress field is not possible in the viscoelastic regime, 
this traction is transmitted primarily through an isotropic pressure perturbation in the solvent pores (i.e., the pore-pressure) of the gel. 
Matching this perturbation to the far-field pressure necessarily generates a nonzero pressure gradient, 
which by Darcy's law drives solvent flow and hence coarsening.

However, at sufficiently long times, $t>\tau_{\mathrm{el}}$, the local and transient \textit{viscoelastic} polymer relaxation modes
coupled to the coarsening-induced stress have relaxed, and the polymer network responds collectively and elastically
\cite{Ferry,RubinsteinColby}. Although the elastic modulus is a property of the network at all times, the gel stress is only \emph{spatially long-ranged and temporally persistent}  in this elastic regime.  In this regime, the Young--Laplace interfacial \emph{traction}
(i.e., the normal force per unit area---equivalently, the normal component of the stress---transmitted across
the interface) has a characteristic magnitude set by the Laplace-pressure scale, $\delta p \sim \gamma/\lambda$, at the
solvent--gel interface. As shown in Appendix \ref{app:spherical_inclusion_elastic_bc}, boundary traction is then balanced by the long-ranged \textit{deviatoric} (non-isotropic) stress field
established in the gel, but the pressure gradient is zero.

Crucially, the resulting pore-pressure field  is \emph{not assumed} to have zero gradient: it is determined by solving the elastic
equilibrium problem in the surrounding gel. Writing the gel stress
as $\boldsymbol{\sigma}=-p\,\mathbf{I}+2G\boldsymbol{\varepsilon}$ and imposing spherical mechanical equilibrium, the elastic
displacement/strain field that balances the interfacial traction yields $dp/dr=0$, i.e.\ the isotropic pressure in the solvent-filled pores of the gel is spatially uniform (with the constant fixed by the far-field boundary condition) (see Appendix \ref{app:spherical_inclusion_elastic_bc}). Therefore, in this elastic regime the
pressure gradient entering Darcy's law vanishes, so the pressure-gradient mechanism for solvent transport is eliminated.
By Darcy's law, this removes the driving force for solvent flow and thus halts the capillarity-driven coarsening.
We therefore refer to \emph{elastic arrest} as the regime in which the interfacial capillary traction is balanced by the gel's
elastic \textit{shear} stress, leaving no pore-pressure gradient to drive further Darcy permeation.

It is also important to note that the elastic stress at the domain boundary is \emph{not fixed a-priori}.
It must be determined by solving the elastic displacement problem in the gel surrounding
the \emph{entire} domain. 
For a spherical domain embedded in an infinite elastic gel \cite{LandauLifshitz}, 
the radial displacement field decays as (see Appendix \ref{app:spherical_inclusion_elastic_bc})
\begin{equation}
u(r) \sim \delta \lambda \left(\frac{\lambda}{r}\right)^2 ,
\end{equation}
so that the normal stress at the interface is proportional to the 
interfacial strain,
\begin{equation}
\sigma_{\rm el} \sim G \frac{\delta \lambda}{\lambda},
\end{equation}
up to a numerical prefactor.
Physically, $\delta \lambda$ represents the instantaneous radial deformation of the gel at the domain boundary induced by the capillary pressure acting on the curved interface. 
It is not an imposed equilibrium mismatch between two stress-bearing phases; rather, it arises dynamically from the requirement that the gel satisfy mechanical equilibrium with the instantaneous Laplace pressure
$\delta p(t)\sim\gamma/\lambda(t)$.
In the elastic regime it is defined at all times as the solution of the elastic equilibrium equations with the \emph{instantaneous interfacial pressure} applied as a boundary condition, which in general is time dependent because $\lambda(t)$ evolves during coarsening.

At long times, once the elastic stress has equilibrated spatially, the Laplace force balance at the domain boundary fixes $\delta \lambda$ by equating the elastic stress to the capillary pressure, $G(\delta \lambda/\lambda)\sim\gamma/\lambda$, which yields $\delta \lambda \sim\gamma/G$; thus, although the domain size $\lambda$ is mesoscopic, the interfacial displacement itself is set by a microscopic elastic length of order the gel mesh size. Thus, it is not that the local stress that is smaller than the Laplace pressure, but rather that their balance gives a very small deformation of the micron scale drop in the elastic regime of the gel.  However, it is this very balance that dictates that the solvent pressure gradient is zero (as explained in Appendix D) since the boundary condition of force balance at the solvent domain interface is established by the balance of the elastic stress and capillary pressure, with no further role for a solvent pressure drop to drive Darcy flow in the gel, and hence coarsening.  That is the significance of elastic arrest which is not a free energy condition, but a force balance at the solvent-rich domain interface. Importantly, this boundary force balance fixes the elastic displacement $\delta \lambda$ but it does not by itself determine the arrested domain size itself $\lambda_{\rm arrest}$ which is determined kinetically as discussed below.

To summarize: In the elastic regime of the gel there is only a deviatoric (non-isotropic) stress induced by the capillary traction at the boundary of the solvent-rich domain and  the pressure gradient is zero (see Appendix \ref{app:spherical_inclusion_elastic_bc}). 
Since $\dot\lambda\sim-(k_0/\eta_D)\nabla p$, the solvent flux ceases and coarsening stops.
By contrast, in the viscoelastic coarsening regime, the long-ranged elastic stress field has not yet developed, so the capillary pressure remains effectively unbalanced, a pressure gradient exists in the gel and by the Darcy flow equation, solvent flow and coarsening proceed.
This does not apply in a coherent $A$--$B$ solid solution (of two elastic phases), since transport is diffusive and driven by gradients of \textit{chemical potential}, which is not proportional to the pressure gradient in an \textit{elastic} system.  The chemical potential  contains contributions from the \emph{entire} elastic stress field (both isotropic pressure and deviatoric, non-isotropic stress).  Consequently, in the gel elastic arrest of the solvent-rich, \textit{fluid} domains occurs when $\nabla p=0$, whereas in a \textit{solid} solution elasticity modifies the thermodynamic driving force itself rather than eliminating transport through the vanishing of a pressure gradient.

\begin{figure}
    \centering
    \includegraphics[width=1\linewidth]{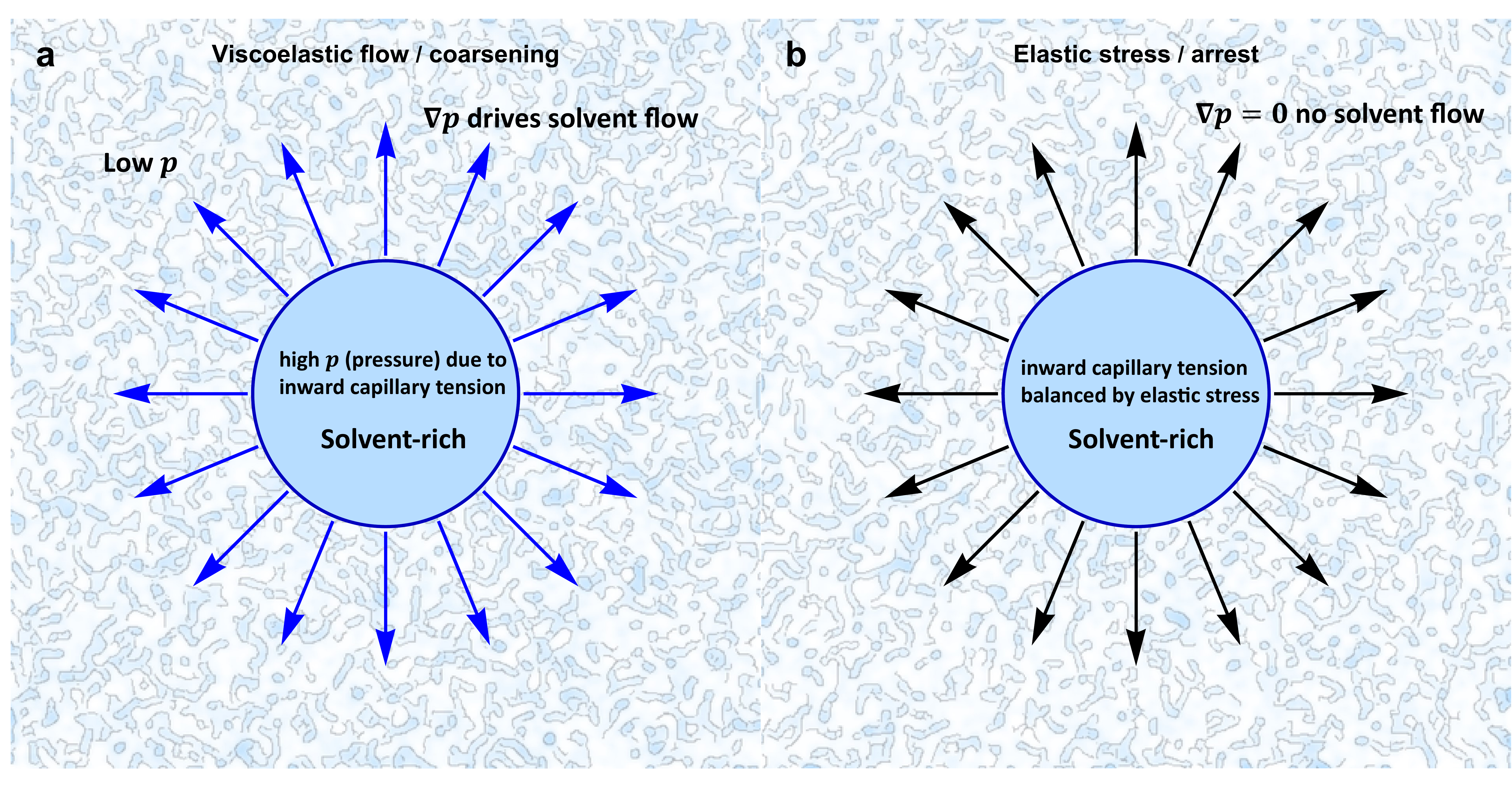}
    \caption{Left: Viscoelastic regime. A solvent-rich domain is at higher pressure due to capillarity. Because elastic stresses in the gel (white – polymers, blue – solvent)  relax rapidly in this regime compared to the coarsening time, the interface condition is primarily a balance between capillary stress and fluid pressure. This generates a pressure difference between the domain and the surrounding solvent-filled pores, leading to a pressure gradient in the pore fluid, Darcy flow, and coarsening.
Right: Elastic regime. The polymer network sustains elastic stress, which balances the capillary stress at the domain boundary. Mechanical equilibrium is then achieved without a pressure difference between the domain and the surrounding pore fluid. Because the same solvent phase occupies both regions, this implies a uniform pressure ($\nabla p=0$ ), eliminating Darcy flow and arresting coarsening.}

    \label{fig:Schematic6}
\end{figure}

We emphasize that elastic arrest is a kinetic effect that is relevant only for $t>\tau_{\mathrm{el}}$. Arrest therefore occurs when the coarsening time required to reach a domain size $\lambda$ becomes comparable to the intrinsic elastic crossover time:
\[
t(\lambda_{\rm arrest}) \sim \tau_{\mathrm{el}}.
\]
Using the slow-poroelastic time-to-size relation of Eq.~\ref{eq:onequarter} ($\lambda^4\propto t$), we obtain the arrest size in the slow-diffusion regime,
\begin{equation}
\lambda_{\rm arrest}^{(\rm slow)} \sim
\Bigg(\,\frac{\eta_{ps}\,\eta_{el}}{\eta_D^2}\,
\frac{k_0\,\gamma^2}{G^2}\,\Bigg)^{1/4},
\end{equation}
where we have expressed the elastic crossover time in terms of an effective elastic viscosity $\eta_{el}=G\tau_{\mathrm{el}}$, because viscosities scale as a modulus times a characteristic relaxation time.  This form highlights that, in contrast to the coarsening kinetics, the arrest length depends only on ratios of viscosities and on elastic and interfacial material parameters.

In the fast-poroelastic regime, the corresponding arrest size is
\begin{equation}
\lambda_{\rm arrest}^{(\rm fast)} \sim
\Bigg(\,\frac{\eta_{el}}{\eta_D}\,
\frac{k_0\,\gamma}{G}\,\Bigg)^{1/3}.
\end{equation}

\section{Results}
In this section, we analyze the previous results to predict the scaling of the domain coarsening as a function of time and gel elastic modulus, as well as the modulus dependence of the arrest length. In addition, we show that this length can be magnified to a value far larger than the gel mesh size, by the viscosity ratios defined above.  We present our analysis for two limiting cases of the dependence of the gel properties (permeability $k_0$ and interfacial tension with the solvent-rich phase $\gamma$) that enter the capillary-driven coarsening and eventual arrest. We keep the explicit dependence on the gel modulus $G$ since this is often varied and measured in experiments \cite{FernandezRico2024NatMat}. The first scenario we call the mesh-scaling gel where the polymer fraction in the gel is relatively small and the mesh size $\xi$ is the largest length in the homogeneous gel.  The permeability in the Darcy equation has units of area and hence scales as $\xi^2$, while the tension (in units of $k_BT$) scales as $1/\xi^2$.  The second scenario we call the melt-like gel where the polymer fraction in the gel is relatively large (order 1/2) and the mesh size is nanometric and can be of order the polymer persistence length, solvent molecule size and other microscopic lengths. In contrast to the mesh-scaling gel with large pores which are the channels for solvent flow, the melt-like gel may be conceptualized as somewhat swollen and crosslinked polymer melt. This is the limit that is applicable to the gels in \cite{FernandezRico2024NatMat} where the arrest length has been measured. For the melt-like gel, we keep $k_0$ and $\gamma$ as parameters which vary slowly with the gel mesh size, due to the involvement of the microscopic length scales of the system. 

We first present the scaling relations for the coarsening dynamics and the arrest length as functions of the three viscosities defined above: The Darcy viscosity $\eta_D$, the poroelastic viscosity for stress diffusion in the \textit{viscoelastic} regime of the polymers $\eta_{ps}$, and the viscosity $\eta_{el}= G \ \tau_{el}$ related to the time scale $\tau_{el}$, at which the viscoelastic regime (present at short times even in a crosslinked gel) crosses over to the elastic regime.  Afterwards, we rewrite the scaling relations as functions of time scales associated with the Darcy and poroelastic viscosities, guided by the fact that a viscosity is the product of a modulus and a time scale. This 
 effective modulus characterizes the short-time resistance of the polymer network to deformation by solvent flow.  For scaling purposes, the distinction between this viscoelastic modulus and the long-time elastic modulus $G$ affects only prefactors and is therefore neglected since both are governed by the same crosslinked, network structure \cite{RubinsteinColby, Ferry}. We therefore write $\eta_D=G \ \tau_D$ and $\eta_{ps}=G \ \tau_{ps}$ where the dependence on $G$ reflects that the solvent flow, is impeded by friction with the gel even in its \textit{viscoelastic} regime.  The results for the scaling relations are shown in Tables \ref{tab:prefactors_cases12} and \ref{tab:prefactors_cases34}. 

 \begin{table}[ht]
\caption{Coarsening laws and kinetic-arrest lengths with dimensional prefactors
in the fast and slow poroelastic regimes, for the two gel models with independent
viscosities.}
\label{tab:prefactors_cases12}
\footnotesize
\begin{ruledtabular}
\begin{tabular}{llll}
\parbox[t]{0.30\linewidth}{\raggedright Gel Model}
&
\parbox[t]{0.10\linewidth}{\raggedright Regime}
&
\parbox[t]{0.30\linewidth}{\raggedright Coarsening law $\lambda(t)$}
&
\parbox[t]{0.30\linewidth}{\raggedright Arrest length $\lambda_{\rm arrest}$}
\\

\parbox[t]{0.30\linewidth}{\raggedright
\textbf{(i) Mesh-scaling gel:}
$G\sim k_BT/\xi^3$,
$k_0\sim\xi^2$,
$\gamma\sim k_BT/\xi^2$}
&
Fast
&
$\displaystyle
\lambda(t)\sim
\left(\frac{k_BT}{\eta_D}\,t\right)^{1/3}
$
&
$\displaystyle
\lambda_{\rm arrest}^{(\mathrm{fast})}\sim
\left(\frac{k_BT}{G}\right)^{1/3}
\left(\frac{\eta_{\rm el}}{\eta_D}\right)^{1/3}
$
\\[2pt]

&
Slow
&
$\displaystyle
\lambda(t)\sim
(k_BT)^{1/3}G^{-1/12}
\left(\frac{\eta_{ps}}{\eta_D^2}\right)^{1/4}t^{1/4}
$
&
$\displaystyle
\lambda_{\rm arrest}^{(\mathrm{slow})}\sim
\left(\frac{k_BT}{G}\right)^{1/3}
\left(\frac{\eta_{ps}\eta_{\rm el}}{\eta_D^2}\right)^{1/4}
$
\\[4pt]

\parbox[t]{0.30\linewidth}{\raggedright
\textbf{(ii) Melt-like gel:}
$\gamma,k_0$ independent}
&
Fast
&
$\displaystyle
\lambda(t)\sim
\left(\frac{k_0\gamma}{\eta_D}\,t\right)^{1/3}
$
&
$\displaystyle
\lambda_{\rm arrest}^{(\mathrm{fast})}\sim
\left(\frac{k_0\gamma}{G}\,
\frac{\eta_{\rm el}}{\eta_D}\right)^{1/3}
$
\\[2pt]

&
Slow
&
$\displaystyle
\lambda(t)\sim
\left(\frac{k_0\gamma^2}{G}\,
\frac{\eta_{ps}}{\eta_D^2}\,t\right)^{1/4}
$
&
$\displaystyle
\lambda_{\rm arrest}^{(\mathrm{slow})}\sim
\left(\frac{k_0\gamma^2}{G^2}\,
\frac{\eta_{ps}\eta_{\rm el}}{\eta_D^2}\right)^{1/4}
$
\\

\end{tabular}
\end{ruledtabular}
\end{table}

\begin{table}[ht]
\caption{Same as Table~\ref{tab:prefactors_cases12}, but rewriting all viscosities
as products of an elastic modulus and characteristic time scales.}
\label{tab:prefactors_cases34}
\footnotesize
\begin{ruledtabular}
\begin{tabular}{llll}
\parbox[t]{0.30\linewidth}{\raggedright Gel Model}
&
\parbox[t]{0.10\linewidth}{\raggedright Regime}
&
\parbox[t]{0.30\linewidth}{\raggedright Coarsening law $\lambda(t)$}
&
\parbox[t]{0.30\linewidth}{\raggedright Arrest length $\lambda_{\rm arrest}$}
\\

\parbox[t]{0.30\linewidth}{\raggedright
\textbf{(i) Mesh-scaling gel + $\eta\sim G\tau$:}
$G\sim k_BT/\xi^3$,
$k_0\sim\xi^2$,
$\gamma\sim k_BT/\xi^2$}
&
Fast
&
$\displaystyle
\lambda(t)\sim
(k_BT)^{1/3}G^{-1/3}
\left(\frac{t}{\tau_D}\right)^{1/3}
$
&
$\displaystyle
\lambda_{\rm arrest}^{(\mathrm{fast})}\sim
\left(\frac{k_BT}{G}\right)^{1/3}
\left(\frac{\tau_{el}}{\tau_D}\right)^{1/3}
$
\\[2pt]

&
Slow
&
$\displaystyle
\lambda(t)\sim
(k_BT)^{1/3}G^{-1/3}
\left(\frac{\tau_{ps}}{\tau_D^2}\,t\right)^{1/4}
$
&
$\displaystyle
\lambda_{\rm arrest}^{(\mathrm{slow})}\sim
\left(\frac{k_BT}{G}\right)^{1/3}
\left(\frac{\tau_{ps}\tau_{el}}{\tau_D^2}\right)^{1/4}
$
\\[4pt]

\parbox[t]{0.30\linewidth}{\raggedright
\textbf{(ii) Melt-like gel + $\eta\sim G\tau$:}
$\gamma,k_0$ independent}
&
Fast
&
$\displaystyle
\lambda(t)\sim
\left(\frac{k_0\gamma}{G}\,
\frac{t}{\tau_D}\right)^{1/3}
$
&
$\displaystyle
\lambda_{\rm arrest}^{(\mathrm{fast})}\sim
\left(\frac{k_0\gamma}{G}\,
\frac{\tau_{el}}{\tau_D}\right)^{1/3}
$
\\[2pt]

&
Slow
&
$\displaystyle
\lambda(t)\sim
\left(\frac{k_0\gamma^2}{G^2}\,
\frac{\tau_{ps}}{\tau_D^2}\,t\right)^{1/4}
$
&
$\displaystyle
\lambda_{\rm arrest}^{(\mathrm{slow})}\sim
\left(\frac{k_0\gamma^2}{G^2}\,
\frac{\tau_{ps}\tau_{el}}{\tau_D^2}\right)^{1/4}
$
\\

\end{tabular}
\end{ruledtabular}
\end{table}

We note that in Table \ref{tab:prefactors_cases34} where we have extracted the maximal dependence on the elastic modulus $G$ that is physically reasonable, the fast and slow coarsening for the mesh-scaling gel both scale as $1/{G}^{1/3}$ with respective time dependence of $t^{1/3}$ and $t^{1/4}$ respectively.  For the melt-like gel, appropriate for high polymer volume fractions where the porosity and interfacial tension depend only weakly on $G$, the fast coarsening law exhibits the same scaling as for the mesh-scaling gel. On the other hand, the slow coarsening law still scales as $t^{1/4}$ but with a coefficient that now scales as $1/\sqrt{G}$.   This is also the modulus scaling of the  arrest length in the slow poroelastic limit. 

We now discuss numerical estimates of the magnification factors related to the viscosity (or time scale) ratios in Tables 1 and 2. Ref.~\cite{FernandezRico2024NatMat} does not present rheological data for their samples so we use two other experimental studies of the rheology of PDMS gels \cite{TakahashiPDMSGel2006, Petet2021} to estimate the relevant quantities for the solvent free case. The solvent  viscosity $\eta_D$ applicable to Ref.~\cite{FernandezRico2024NatMat} is that of the small molecule, organic solvent, HFBMA (heptafluorobutyl methacrylate)  \cite{FernandezRico2024NatMat}.  Its molecular weight is $\sim 300$~g/mol but to the best of our knowledge, its viscosity is not found in the literature.  The viscosity of a similar solvent, BMA (n-butyl methacrylate) whose molecular weight $\sim 150$~g/mol, is known and has the value $\sim 10^{-3}$~Pa$\cdot$~s \cite{RoehmBMA} which is what we estimate for $\eta_D$.   

The gels in \cite{TakahashiPDMSGel2006} are more sparsely crosslinked than those of Ref.~\cite{FernandezRico2024NatMat} which may overestimate the polymeric viscosities while the polymers in \cite{Petet2021} are longer than those of Ref.~\cite{FernandezRico2024NatMat} which may also overestimate the viscosities.  For the same reason, both may overestimate the time for crossover from viscoelastic to elastic response $\tau_{el}$. Both of those papers suggest an estimate of the \textit{local} $\eta_{ps} \ > 10 \text{Pa} \cdot \text{s}$ for viscosity of the gel in its viscoelastic regime. We use the value of $\eta_{ps} \sim 100 \text{Pa} \cdot \text{s}$  obtained from the value of $G''$ at the crossover $G''=G'$ in Fig. 3 (for the solvent-free gel 1E) of  Ref. \cite{TakahashiPDMSGel2006}. We note that the viscosity is proportional to $G''/\omega$ which increases even more as one goes further into the viscoelastic regime \cite{Chambon1985}. In addition, the true coarse-grained \textit{transport} viscosity governing solvent redistribution relevant to our model of gel response to solvent flow, may be substantially larger.  The estimated \cite{TakahashiPDMSGel2006, Petet2021}    frequency of $\omega \sim 100/{\text{s}}$ for the crossover from viscoelastic to elastic, estimated from where $G'=G'"$ (Fig. 3 of \cite{TakahashiPDMSGel2006} for gel 1-E)  corresponds to a time of $\tau_{el} \sim$~0.01s inferred from local rheological measurements.  Again, this time scale may be larger in the presence of solvent redistribution in the porous network.  The latter predicts a range of $\eta_{el}$ for the gels of Ref.~\cite{FernandezRico2024NatMat} where the Young's modulus ranges from 10--100~kPa, with an average value of $\eta_{el} \sim 5 \cdot 10^2 \text{Pa} \cdot \text{s}$.  Thus the amplification factors in Table 1 (for the non-dilute limit relevant to \cite{FernandezRico2024NatMat} ) are approximately $\eta_{ps}/\eta_D \sim 10^5 $ and $\eta_{ps} \eta_{el}/\eta_D^2  \sim 10^7$.  Our estimate of $\eta_{ps}/\eta_D \sim 10^5 $ gives a value for the crossover length of Eq.~\ref{eq:xover} of the order of microns (ignoring numerical prefactors), since the ratio $\gamma/G$ is of the order of the nanometric mesh size.  This means that the system is in the slow, poroelastic regime of Tables 1 and 2.   Even the 4th root of this ratio in the expression for $\lambda_{\rm arrest}^{(\mathrm{slow})}$ still gives a magnification of the nm scale associated with the mesh size by a factor of $10^{7/4} \sim 50$.  Finally, we note that our estimated value of $\tau_{el} \sim 0.01s $ suggests that elastic arrest occurs at short times after the quench.  This may explain the observation in \cite{FernandezRico2024NatMat} that micron-scale solvent and gel regions are already established after the quench when images are taken at a frame rate of 0.1 frames/sec. 

We reiterate that the viscosities defined here $\eta_{ps}$ and $\eta_{el}$ refer to the gel viscosity in the presence of Darcy flow of the solvent.  The rheological measurements referred to in previous paragraph only provide the local effects of the polymer rearrangements.  However, in the presence of solvent flow, there may be potentially substantially larger viscosities and crossover times that account for the effects of hydrodynamic  and pore-size constraints on the polymers during solvent flow. All these considerations highlight the importance of correlating structural and rheological properties of phase separating gels and should motivate rheological measurements of these gels \cite{FernandezRico2024NatMat}  with and without Darcy flow of solvent.

 \section{\label{sec:discussion}Discussion}
\textit{Interpretation of mesh-scaling gel: scaling of domain coarsening with time and elastic modulus:} The only length scale in the problem is the gel mesh size $\xi \sim 1/G^{1/3}$.   The fast poroelastic diffusion predicts coarsening which scales as $t^{1/3}$ while slow poroelastic diffusion coarsening scales as $t^{1/4}$.  If we further take the viscosities to all scale as the modulus $G$ and a different effective relaxation times as in Table \ref{tab:prefactors_cases34}, the dependence on the elastic modulus in all cases, for both the coarsening law and kinetic arrest lengths scales as the mesh size, proportional to $1/G^{1/3}$. The coarsening lengths which are power laws in the time, naturally contain a characteristic time scale.  For the fast coarsening this is the relaxation time of the Darcy flow $\tau_D$ appropriate for solvent flow against the friction of the polymers in their static configurations. The slow coarsening which proceeds via poroelastic stress diffusion in the gel with the flowing solvent, has a characteristic time proportional to the ratio $\tau_D^2/\tau_{ps}$. Since $\tau_{ps}$ is related to the local and transient viscoelastic relaxation of the polymer chains induced by the solvent flow, in systems with a high polymer fraction, we expect $\tau_{ps}\gg \tau_D$ so the characteristic time scale of the slow coarsening is small compared to $\tau_D$ itself. The fact $\tau_{ps}\gg \tau_D$ is the reason that the crossover length scale $\lambda_\times$ in Eq.~\ref{eq:xover} is magnified from the mesh scale to the mesoscale.  Nevertheless, the ``fast'' and ``slow'' nomenclature refers to the
coarsening exponents rather than the characteristic time scales:
even when the slow poroelastic mode is governed by a shorter characteristic
time scale, it still yields slower growth in time ($t^{1/4}$ versus $t^{1/3}$).

\textit{Melt-like gel: scaling of domain coarsening with time and elastic modulus:} For this system, we keep the porosity and interfacial tension as parameters that vary only slowly with the modulus $G$, since for large polymer fraction gels (the melt-like limit), additional length scales related to the Kuhn length and solvent size can determine these parameters.  However, despite this, the crosslinked, melt-like gel is still well characterized by its elastic modulus $G$ which remains the dominant  parameter controlling both stress transmission and  arrest.   In this regime,  fast poroelastic diffusion predicts coarsening which scales as $t^{1/3}$ while slow poroelastic diffusion coarsening scales as $t^{1/4}$.  If we further take the viscosities to all scale as the product of the gel modulus $G$ and  different effective relaxation times as in Table \ref{tab:prefactors_cases34}, for fast poroelastic coarsening, the dependence on the elastic modulus in all cases, for both the coarsening  and the  arrest lengths scale with  $1/G^{1/3}$. 

On the other hand, for slow poroelastic coarsening of the melt-like gel, we stress that both the coarsening and  arrest lengths of Table \ref{tab:prefactors_cases34} scale as $1/\sqrt{G}$. The arrest length is  magnified from its microscopic or mesh-size value by the ratio of the time scales $\tau_{el}\gg \tau_D$ for fast coarsening and $\tau_{el} \tau_{ps} \gg \tau_D^2$ for slow coarsening.  
These large ratios reflect the separation between solvent transport and polymer stress relaxation processes:
 $\tau_{ps}$ is related to the local and transient viscoelastic relaxation of the polymer chains activated by the solvent flow, while $\tau_D$ is related to solvent flow in the presence of polymer friction in an essentially static configuration. The longest time scale is $\tau_{el}$ which is the crossover from viscoelasticity (activated by solvent flow) to elastic response when further coarsening is kinetically arrested by elastic stress acting over a region of order $\lambda$. We emphasize that the arrest discussed here refers to the breakdown of capillarity-driven, pressure-mediated coarsening; while other, much slower mechanisms of structural evolution cannot be excluded in principle, they lie outside the scope of the present theory.

\textit{Comparison with experiment:} The melt-like gel predictions may be applicable to the experiments of \cite{FernandezRico2024NatMat} where the gels are covalently crosslinked and the polymer fraction is relatively large (of order 1/2). Unfortunately, the coarsening kinetics has not yet been measured in these systems.   Our predictions for slow poroelastic diffusion for the melt-like gel, for scaling of the coarsening length with $t^{1/4}$ and with elastic modulus as $1/\sqrt{G}$, do agree with the numerical model calculations in \cite{FernandezRico2024NatMat} but the significance of this agreement is not clear since their model is based on Cahn-Hilliard dynamics with a constant mobility while our scaling predictions originate in the Darcy driven solvent flow.   Although a $t^{1/4}$ growth law is also known \cite{Bray1994} to arise in coarsening via 
surface–diffusion–limited coarsening where material transport occurs
along connected interfaces, this mechanism is not applicable to the present
system since observability of coarsening in the experiments (see the SI of \cite{FernandezRico2024NatMat} ) depends on the elastic modulus of the bulk material. In addition, in the three softest gels of \cite{FernandezRico2024NatMat} the solvent-rich domains are distinct and unconnected.

The primary prediction of our scaling model that does agree with the measurements in \cite{FernandezRico2024NatMat} is that the arrest length-scale varies for the melt-like gel for slow poroelastic diffusion, as $1/G^{1/2}$ (see Appendix~\ref{app:lambdastar}). This agreement indicates that the gels in these polymer-rich systems are most probably not in the mesh-scaling  or the fast poroelastic diffusion regimes,  where the arrest length scale is predicted to vary as $1/G^{1/3}$.  Our scaling relations also predict the dependence of the arrest-length on the various viscosities and future measurements of this behavior (see below) would be of interest.

We now address the very slow drainage of solvent at time scales of many hours. That process, is also driven by capillarity involving the solvent in the gel \textit{near its boundary} with an excess solvent \textit{reservoir}. The solvent flow is slow and we expect that the  drainage  initiated at the sample boundaries (not included in the other sections of this paper) will depend on poroelastic diffusion in that region, with the  Darcy flow near the boundary related to the solvent viscosity $\eta_s$ and not $\eta_{ps}$; this is because the polymers have relaxed all stresses on the macroscopic time scale. We thus estimate that the macroscopic time for drainage at the boundary scales as $R^2/D_D$ where $R$ is a macroscopic  size and the Darcy poroelastic diffusion constant $D_D$ scales as $k_0 G/\eta_s$.  For an order of magnitude numerical estimate, we take the permeability $k_0\sim \xi^2 \sim (k_B\ T/G)^{2/3}$ and $\eta_S$ to be of the order of magnitude of the water viscosity, $10^{-3} \text{Pa~s}$. For $G \sim 100 \text{kPa}$ as a typical rigidity \cite{FernandezRico2024NatMat}, we find for a 1cm sample, a drainage time  of the order of one day.

 We conclude by emphasizing that the scope of our scaling picture is more general than its application to specific gels of Ref.~\cite{FernandezRico2024NatMat}, which we interpret as
representative of melt-like gels.
Our predictions for the mesh-scaling gels can be tested by
working at significantly lower polymer volume fractions, where the gel
enters the mesh-scaling regime and distinct coarsening laws and arrest
scales are expected.

Even more generally, our framework applies to both permanently and
non-permanently crosslinked gels.
In non-permanently crosslinked gels, where the homogeneous swollen state
already exhibits viscoelastic stress relaxation, the effective viscosities
introduced here can be directly related to the linear rheological
properties of the uniform gel.
However for permanently crosslinked gels, the viscosities appearing in our theory are intrinsically
non-equilibrium quantities, controlled by polymer chain relaxation in the
presence of solvent \emph{flow}, such as that induced by shrinking
solvent-rich domains.
In such systems, these effective viscosities may be independently probed
through measurements of the kinetics of solvent uptake or release of a
macroscopic gel in contact with a solvent reservoir.
This allows experimental tests of the dynamical
ingredients of our theory across a wide range of gel architectures.  Such measurements (and even more standard rheological ones) would allow experimentalists to correlate gel dynamics and structure in phase separating systems.

\section{\label{sec:acknowledgments}Acknowledgements}

We are grateful to Eric Dufresne and Carla Fernandez-Rico for discussions about their experiments and to Ram Adar, David Andelman, Haim Diamant, Amit Kumar, Biplab Mandal, Manu Mannattil, and  David Zwicker for  insights into their theoretical models.

\appendix

\section{Curk and Luijten model}
\label{app:curk}
A paper with theory and model calculations of phase separation and coarsening in a viscoelastic gel by Curk and Luijten \cite{CurkLuijten2023PNAS} extended LSW-type Ostwald ripening driven by capillarity (interfacial tension $\gamma$) by incorporating elastic stresses and their viscoelastic relaxation in the surrounding matrix. Their approach is based on diffusion-controlled evaporation–condensation through a viscoelastic solid, modified by elastic and viscoelastic effects, and does not include poroelastic (Darcy) solvent flow through a deformable, porous network, which provides the dominant dissipative mechanism for relaxing capillary pressure in our approach (see Appendix for further discussion). They predict an initial regime of growth followed by elastic arrest at short times, an intermediate viscoelastic coarsening regime with linear-in-time growth, and a late-time crossover back to LSW coarsening with domain size scaling as $t^{1/3}$. The late-time return to LSW scaling is not observed in PDMS gels, where the long-time (hours) state is one of elastic arrest, indicating a distinct rate-limiting physical mechanism.

\section{Onuki-Puri numerical calculations}
\label{app:onukipuri}
A previous, numerical calculation by Onuki and Puri \cite{OnukiPuri1999}  focused on  a two-dimensional  crosslinked gel with (implicit) solvent-gel friction. In their numerical solutions the onset of elastic arrest occurred at a time that depended on the value of the modulus. They found that for an uncrosslinked gel (their effective crosslink density, is equivalent to the modulus $G$ in units of $k_BT$) coarsening was governed by a 1/3 power law.  When varying degrees of crosslinking were included, the coarsening was governed by an effective power law where the exponent depended on the crosslink density (see their Fig. 2 and caption).   They also observed the onset of elastic arrest with a time that depended on the value of the modulus.  The dynamics of their model only included polymer-solvent friction and the gel had no viscoelastic regime and could not include  Darcy transport, poroelastic stress diffusion, and viscoelastic stress buildup, all important elements of our approach and its relevance to the experiments in \cite{FernandezRico2024NatMat}.  In their simulations, the effective network--solvent friction includes an explicit
Stokes prefactor (\(6\pi\)), which enters directly into the arrest criterion.
As a result, the arrested length can be enhanced by roughly an order of
magnitude relative to the naive elastocapillary estimate \(\gamma/G\), even in
the absence of an additional large control parameter.
By contrast, in our scaling theory the viscosities enter only through ratios
(see the Tables below), so such numerical prefactors cancel.
The remaining microscopic input enters
the arrest length only to a fractional power, rendering the predicted
mesoscopic magnification robust and weakly sensitive to microscopic details.

\section{Unified diffusion-based description of coarsening in fluids and viscoelastic gels}
\label{app:unified_diffusion}

In this Appendix we present a unified derivation of the coarsening dynamics for
(i) phase separation in a binary fluid mixture of small molecules, and
(ii) solvent-rich domains embedded in a polymer gel.
Although the microscopic transport mechanisms differ in the two cases, both can be formulated in terms of a diffusion equation for the  chemical potential or equivalently, for the pressure; both use the capillary (Laplace pressure jump) boundary condition at the domain interface.
This unified formulation clarifies which approximations underlie the standard LSW result for simple fluids and how these are modified in gels by slow poroelastic transport.

\subsection{Interfacial flux and capillary boundary condition}

We consider a spherical solvent-rich domain of radius $\lambda(t)$ that is shrinking in time.
Let $N$ denote the number of transported molecules (B molecules in a binary fluid, or solvent molecules in a gel) inside the domain.
Mass conservation at the moving interface gives
\begin{equation}
\frac{dN}{dt} = - J S ,
\label{eq:flux_conservation_app}
\end{equation}
where $J$ is the outward flux of transported molecules evaluated at the domain boundary and
$S = 4\pi \lambda^2$ is the surface area of the domain.
The flux is normal to the interface and is controlled by the local driving force acting at the boundary.

To relate $\dot N$ to the interface velocity $\dot\lambda$, we note that a change in radius by $d\lambda$
changes the domain volume by $dV = S\,d\lambda$.
The \emph{net} number of transported molecules lost from the domain when a volume $dV$ is removed is
\(\Delta c\, dV\), where
\begin{equation}
\Delta c \equiv c_{\rm in} - c_{\rm out}
\end{equation}
is the concentration difference between the inside of the domain and the surrounding phase.
Therefore,
\begin{equation}
\dot N \sim \Delta c\, S\, \dot\lambda.
\label{eq:Ndot_Delta_c}
\end{equation}
For a shrinking domain, $\dot\lambda<0$, and hence $\dot N<0$ as expected.
Combining Eqs.~\eqref{eq:flux_conservation_app} and \eqref{eq:Ndot_Delta_c} gives the interfacial evolution law
\begin{equation}
\Delta c\,\dot\lambda \sim - J.
\label{eq:lambda_flux_general}
\end{equation}

The flux can be written in terms of the chemical potential $\mu$ of the transported species as
\begin{equation}
J = - M \nabla \mu ,
\label{eq:flux_mu_app}
\end{equation}
where $M$ is a mobility.

\subsection{Mobility in binary fluids and gels}

The mobility \(M\) appearing in the interfacial flux relation,
\(J = - M \nabla \mu\),
originates in the microscopic mechanism by which the transported species moves in response
to a chemical potential gradient.
Its physical meaning differs in the two systems considered here.

\paragraph{Binary fluid (small molecules).}
For a mixture of small molecules, transport occurs by ordinary molecular diffusion.
The mobility is related to the mutual diffusion constant \(D\) via the Einstein relation,
\begin{equation}
	M \sim D\, c_B ,
\end{equation}
where \(c_B\) is the number density of the transported species.
Equivalently, using \(\nabla \mu = \nabla p / c_B\) which is appropriate for \textit{fluids}, the flux may be written as
\(J \sim - D \nabla c_B\).
In this case, the mobility is due to the rapid diffusive motion of individual molecules
in the surrounding fluid.

\paragraph{Gel (poroelastic transport).}
In a gel, transport of solvent is instead controlled by permeation through a deformable
polymer network.
The appropriate constitutive relation is Darcy’s law,
\begin{equation}
	\vec v = - \frac{k_0}{\eta_D} \nabla p ,
\end{equation}
where \(k_0\) is the permeability of the gel and \(\eta_D\) is the viscosity associated with
solvent flow impeded by friction with the polymer network.
The solvent flux is \(J = c_s v\), where $c_s$ is the solvent concentration, so that the mobility entering the flux relation is
\begin{equation}
	M \sim \frac{k_0\, c_s}{\eta_D} .
\end{equation}
In contrast to the binary-fluid case, the mobility here is controlled by hydrodynamic
permeation of solvent through the network rather than by molecular diffusion.

\medskip
Thus, while the diffusion equation for the pressure field has the same mathematical form
in both systems, the mobility \(M\)—and therefore the physical origin of dissipation—is
fundamentally different in binary fluids and in gels.

In both cases the capillary driving force originates in the curvature dependence of the chemical potential at the interface.
For a spherical domain of radius $\lambda$, the chemical potential at the interface exceeds that of a flat interface by
\begin{equation}
\delta \mu = \frac{\gamma}{c \, \lambda},
\label{eq:capillary_mu_app}
\end{equation}
where $\gamma$ is the interfacial tension and $c$ is the concentration of the transported species in the domain.

\subsection{Relation of mechanical and osmotic pressures}
In an incompressible soft-matter system, local variations in solvent concentration generate osmotic stresses that tend to drive solvent flow. Because local volume changes are forbidden, these stresses give rise to spatial variations of a mechanical pressure field, which enforces incompressibility and appears in the force balance governing solvent motion. This mechanical pressure is a physical stress that adjusts so as to balance all forces acting on the material under the incompressibility constraint.

Gradients of chemical potential are directly related to gradients of osmotic pressure, since the osmotic pressure measures the change in free energy associated with locally adding or removing material. Elastic stresses arising from deformation of the polymer network also contribute to the mechanical pressure, but they do so through the network’s mechanical response to imposed stresses rather than through a direct coupling to composition. As a result, elastic stresses influence the pressure field indirectly, by determining how osmotic stresses are transmitted through the network.

In regimes where the polymer network can relax locally — corresponding to the viscoelastic or flowing regime relevant on the coarsening time scale— elastic stresses adjust rapidly and do not build up over large distances. As a result, the pressure gradients that drive solvent flow are set primarily by gradients of osmotic pressure.

By contrast, at sufficiently long times the network responds elastically over length scales comparable to the domain size, and elastic stresses accumulate over these large scales. In this elastic regime, further solvent transport requires deformation of the entire domain, and the resulting pressure gradients are determined by force balance across the domain rather than by local osmotic gradients alone. This buildup of elastic stress leads to elastic arrest of coarsening (see Appendix \ref{app:spherical_inclusion_elastic_bc}.
Thus, while elastic stresses are essential for determining the arrest of coarsening, on the coarsening time scale itself the polymer network behaves viscoelastically, and elastic stresses relax locally and do not contribute significantly to the pressure gradients that drive flow.

Capillary stresses enter separately as boundary conditions at interfaces: the Laplace law fixes the pressure jump across the interface between the solvent-rich domain and the gel and thereby sets the magnitude of the pressure difference that is the boundary condition on  the pressure field in the surrounding material.

In summary, while the mechanical pressure field in fluids includes osmotic and capillary contributions, gradients of chemical potential may be equivalently expressed in terms of pressure gradients because osmotic stresses provide the fundamental thermodynamic driving force, with transient and local elastic stresses controlling how this driving force is transmitted (via energy dissipation) in the viscoelastic regime of the gel. 

\subsection{Relation between chemical potential and pressure gradients.}
For an incompressible binary system, we consider a free-energy density $f(c)$ that depends on the local concentration $c$ of the transported species (the solvent in a gel, or component $B$ in a small-molecule mixture). The chemical potential is
\begin{equation}
\mu = \frac{df}{dc_B}.
\end{equation}
 
For a system containing $N$ particles in a volume $V$ with concentration 
$c = N/V$ and free-energy density $f(c)$, the total free energy is
\[
F = V\, f(c).
\]
The osmotic pressure is defined as the force per unit area opposing a change in volume at fixed particle number,
\[
\Pi \equiv -\left(\frac{\partial F}{\partial V}\right)_{N}.
\]
Using $dc = -(c/V)\, dV$, one finds
\[
\Pi = c \frac{df}{dc} - f(c).
\]
Identifying the chemical potential as $\mu = df/dc$, this can be written as
\[
\Pi = \mu\, c - f(c).
\]

To make this relation explicit, we differentiate $\Pi$
at fixed temperature. Using the chain rule,
\[
d\Pi = \mu\, dc + c\, d\mu - df.
\]
By definition of the chemical potential,
\[
\mu \equiv \frac{df}{dc},
\]
so that
\[
df = \mu\, dc.
\]
Substituting, the terms proportional to $dc$ cancel, leaving
\[
d\Pi = c\, d\mu.
\]
Therefore, when the free energy depends only on concentration,
\begin{equation}
\nabla \Pi = c\, \nabla \mu.
\label{eq:gradPi_gradmu_app}
\end{equation}

\medskip

\noindent
\textbf{Limitation to fluid phases.} The proportionality above relies on the assumption that
\[
f = f(c)
\]
is a function of concentration alone. In an elastic solid or gel, however, the free-energy density depends on additional independent variables such as the strain $\varepsilon$:
\[
f = f(c,\varepsilon).
\]
In that case,
\[
df = \left(\frac{\partial f}{\partial c}\right)_{\varepsilon} dc
   + \left(\frac{\partial f}{\partial \varepsilon}\right)_{c} d\varepsilon,
\]
and the chemical potential is defined at fixed strain,
\[
\mu = \left(\frac{\partial f}{\partial c}\right)_{\varepsilon}.
\]
Repeating the derivation gives
\[
d\Pi = c\, d\mu
- \left(\frac{\partial f}{\partial \varepsilon}\right)_{c} d\varepsilon.
\]

Thus, in an elastic medium where strain varies spatially, gradients in chemical potential are \emph{not} simply proportional to gradients in pressure. The additional mechanical term couples composition and deformation, so transport cannot in general be written purely in terms of $\nabla \Pi$. The relation $\nabla \Pi = c\, \nabla \mu$ holds only for systems whose thermodynamics is fully described by concentration alone (e.g.\ two-component fluids), but not for elastic systems in which strain is an independent degree of freedom.
In the relation of the osmotic pressure and the chemical potential, $c$ denotes the concentration of the transported species (the B molecules in the AB small-molecule mixture or the solvent molecules in the solvent--gel system) outside the coarsening solvent-rich (or B-rich) domain.

In an incompressible system, the mechanical pressure (outside the solvent-rich domain) that drives transport differs from the osmotic pressure only by a reference constant (and by elastic or capillary contributions treated separately through force balance and (Laplace) boundary conditions at the solvent-rich domain interface). Consequently, pressure gradients may be identified with osmotic pressure gradients, $\nabla p = \nabla \Pi$, which gives the thermodynamic relation
\begin{equation}
\nabla \mu = \frac{1}{c}\, \nabla p.
\label{eq:mu_pressure_relation_app}
\end{equation}
This allows the flux to be written equivalently as
\begin{equation}
J = - \frac{M}{c} \, \nabla p .
\label{eq:flux_pressure_app}
\end{equation}
The capillary boundary condition fixes the mechanical pressure at the domain surface through the Laplace (osmotic) pressure jump $\delta p$,
\begin{equation}
p(r=\lambda,t) = \delta p \equiv \frac{\gamma}{\lambda}.
\label{eq:laplace_bc_app}
\end{equation}

\subsection{Diffusion equation for the pressure field outside the coarsening domain}
\label{sec:pressure_diffusion_derivation_app}

We show now that the mechanical pressure field satisfies a diffusion equation,
\begin{equation}
\frac{\partial p}{\partial t}=D\nabla^2 p,
\qquad r\ge \lambda,
\label{eq:diffusion_general_app}
\end{equation}
This result follows from the combination of (i) the expression for the solvent flux, written in terms of pressure gradients, and (ii) local conservation of solvent.

The outward number flux \(J\) of the transported species is proportional to the driving-force
gradient,
\begin{equation}
J = - M \nabla \mu ,
\label{eq:J_mu_app}
\end{equation}
where \(M\) is the mobility.
Using the incompressible thermodynamic relation \(\nabla\mu=\nabla p/c_B\),
this may be written as
\begin{equation}
J = - \frac{M}{c_B}\,\nabla p .
\label{eq:J_p_app}
\end{equation}

Let \(c(\mathbf r,t)\) be the number density of the transported species in the medium outside the
domain (the A-rich phase for a binary fluid, or the polymer-rich phase for a gel).
Local conservation implies
\begin{equation}
\frac{\partial c}{\partial t} + \nabla\!\cdot J = 0 .
\label{eq:local_conservation_app}
\end{equation}
Substituting Eq.~\eqref{eq:J_mu_app} into Eq.~\eqref{eq:local_conservation_app} gives
\begin{equation}
\frac{\partial c}{\partial t} = \nabla\!\cdot\!\big(M\nabla\mu\big).
\label{eq:diffusion_mu_app}
\end{equation}

For small departures from equilibrium \emph{in the surrounding medium outside the domain},
where the pressure and composition fields adjust diffusively to the interfacial boundary
condition, variations in chemical potential are proportional to composition variations,
\begin{equation}
\delta \mu \simeq \left(\frac{\partial \mu}{\partial c}\right)\delta c ,
\end{equation}
so that Eq.~\eqref{eq:diffusion_mu_app} reduces to an ordinary diffusion equation for \(c\) outside the coarsening domain,
\begin{equation}
\frac{\partial c}{\partial t} \simeq D\,\nabla^2 c,
\qquad
D \equiv M\left(\frac{\partial \mu}{\partial c}\right),
\label{eq:D_general_app}
\end{equation}
where \(D\) is the appropriate diffusivity (molecular or poroelastic, depending on the system).

Using the incompressible thermodynamic relation \(\delta\mu=\delta p/c_B\):
\(\delta p \propto \delta c\), so Eq.~\eqref{eq:D_general_app} is equivalently a diffusion equation
for \(p\) in the region outside the coarsening domain: 
\begin{equation}
\frac{\partial p}{\partial t} = D \nabla^2 p ,
\qquad r \ge \lambda ,
\label{eq:diffusion_general_app2}
\end{equation}
with boundary conditions
\begin{equation}
p(r=\lambda,t)=\delta p, \qquad p(r\to\infty,t)=0 ,
\label{eq:bc_general_app}
\end{equation}
and initial condition $p(r,t=0)=0$.
Thus the diffusive spreading of \(p\) (and the associated length scale \(\ell\sim\sqrt{Dt}\))
follows directly from a gradient-driven flux and local conservation.

In a binary fluid mixture of small molecules, \(M\sim D\,c_B\) and the diffusion equation reduces
to ordinary molecular diffusion of composition (or chemical potential) in the continuous phase.
In a gel, the same mathematical structure arises from poroelastic transport: Darcy’s law together
with force balance and network elasticity leads to a diffusion equation for pressure with
\(D=D_p\sim k_0 G/\eta_{ps}\) (see\cite{TanakaFillmore1979})

The physical meaning of the diffusion constant $D$ differs in the two cases considered here:
\begin{itemize}
\item \textit{Binary fluid (small molecules):}
$D$ is the molecular mutual diffusion constant of the B molecules in the A-rich phase.
\item \textit{Gel (poroelastic transport):}
$D=D_p$ is the poroelastic diffusion constant,
\begin{equation}
D_p \sim \frac{k_0 G}{\eta_{ps}},
\end{equation}
where $k_0$ is the permeability of the gel, $G$ its elastic modulus, and $\eta_{ps}$ the viscosity associated with the viscoelastic response of the polymer--solvent system.
\end{itemize}

\subsection{Exact solution around a spherical domain}

For spherical symmetry, the solution of
Eqs.~\eqref{eq:diffusion_general_app2}--\eqref{eq:bc_general_app} is
\begin{equation}
p(r,t) = \delta p  \  \frac{\lambda}{r}
\mathrm{erfc}\!\left(\frac{r-\lambda}{2\sqrt{D t}}\right),
\qquad r \ge \lambda .
\label{eq:pressure_solution_app}
\end{equation}
The radial pressure gradient evaluated at the domain boundary is
\begin{equation}
\left.\frac{\partial p}{\partial r}\right|_{r=\lambda}
= - \delta p
\left(
\frac{1}{\lambda}
+ \frac{1}{\sqrt{\pi D t}}
\right).
\label{eq:gradp_interface_app}
\end{equation}
where $\delta p \sim \gamma/\lambda$.
This expression shows explicitly how the interfacial pressure gradient evolves in time as the pressure field spreads into the surrounding medium.

\subsection{Binary fluid: fast molecular diffusion (LSW limit)}

For a binary fluid of small molecules, the molecular diffusion time across a domain,
$t_D \sim \lambda^2/D$, is much shorter than the coarsening time.
The pressure field therefore rapidly reaches its long-time limit, equivalent to taking $t\to\infty$ in
Eq.~\eqref{eq:gradp_interface_app}.
In this limit,
\begin{equation}
\left.\frac{\partial p}{\partial r}\right|_{r=\lambda}
\simeq - \frac{\delta p}{\lambda}.
\label{eq:gradp_LSW_app}
\end{equation}
Equivalently, the pressure satisfies Laplace’s equation outside the domain.
Substituting Eq.~\eqref{eq:gradp_LSW_app} into the interfacial flux condition (Eqs.~\ref{eq:lambda_flux_general} and \ref{eq:J_p_app}) yields the standard LSW scaling for coarsening in binary fluids.

\subsection{Gel: poroelastic diffusion and crossover regimes}

In a gel, poroelastic diffusion is slow and the time-dependent term in
Eq.~\eqref{eq:gradp_interface_app} must be retained.
Defining the poroelastic diffusion length
\begin{equation}
\ell_p(t)=\sqrt{D_p t},
\end{equation}
the interfacial pressure gradient can be written as
\begin{equation}
\left.\nabla p\right|_{r=\lambda}
= - \delta p
\left(
\frac{1}{\lambda}
+ \frac{1}{\sqrt{\pi}\,\ell_p}
\right).
\end{equation}
This expression exhibits two limiting regimes:
\begin{enumerate}
\item \textit{Slow poroelastic diffusion} ($\ell_p \ll \lambda$):  
\[
\left.\nabla p\right|_{r=\lambda}
\sim - \frac{\delta p}{\ell_p},
\]
corresponding to a transient buildup of the interfacial pressure gradient.
\item \textit{Fast poroelastic diffusion} ($\ell_p \gg \lambda$):  
\[
\left.\nabla p\right|_{r=\lambda}
\sim - \frac{\delta p}{\lambda},
\]
where the interfacial gradient has saturated and no longer depends on time.
\end{enumerate}
Substituting the interfacial pressure gradients derived above into the interfacial flux condition,
Eqs.~\eqref{eq:lambda_flux_general} and \eqref{eq:flux_pressure_app}, yields a closed equation
for the time evolution of the domain size \(\lambda(t)\).
At the scaling level this equation takes the form
\begin{equation}
	\dot{\lambda} \sim - \frac{1}{\Delta c}\,\frac{M}{c_B}\,
	\left.\nabla p\right|_{r=\lambda},
\label{eq:lambda_evolution_scaling_app}
\end{equation}
so that the coarsening kinetics are entirely controlled by the pressure gradient evaluated
at the domain boundary.

\subsection{Coarsening laws}

Inserting the  mobility \(M\) into the interfacial evolution law
\(\Delta c\,\dot{\lambda}\sim- M\,\left.\nabla\mu\right|_{r=\lambda}\)
(or equivalently \(\Delta c\,\dot{\lambda}\sim- (M/c_B)\,\left.\nabla p\right|_{r=\lambda}\))
immediately yields the coarsening kinetics, because the  dynamics are controlled by
the local interfacial gradient evaluated at \(r=\lambda\).

\subsubsection{Binary fluid (small molecules; fast molecular diffusion / LSW).}
For a binary fluid of small molecules, transport in the continuous phase is due to
ordinary molecular diffusion with diffusion constant \(D\).
As shown above, the pressure field outside a shrinking domain satisfies a diffusion equation
and, because molecular diffusion is fast compared to the coarsening dynamics, the pressure
profile rapidly reaches its long-time limit.
Equivalently, the pressure satisfies Laplace’s equation outside the domain, with the
capillary boundary condition \(p(r=\lambda)=\delta p\sim\gamma/\lambda\).

In this limit, the interfacial pressure gradient is time independent and scales as
\[
\left.\nabla p\right|_{r=\lambda}\sim \frac{\delta p}{\lambda}\sim \frac{\gamma}{\lambda^2}.
\]
Using the interfacial evolution law
\(\Delta c\,\dot{\lambda}\sim-(M/c)\left.\nabla p\right|_{r=\lambda}\)
with \(M\sim D c\) for molecular diffusion then gives
\begin{equation}
	\dot{\lambda}\sim
	- \frac{1}{\Delta c}\,\frac{M}{c}\left.\nabla p\right|_{r=\lambda}
	\sim
	- \frac{1}{\Delta c}\,D\,\frac{\gamma}{\lambda^2}.
	\label{eq:lambda_dot_LSW_app}
\end{equation}
Integrating this equation yields the standard LSW coarsening law
\(\lambda(t)\sim (D \, \gamma\,t/\Delta c)^{1/3}\).

\subsubsection{Gel (poroelastic; slow poroelastic diffusion)}
The interfacial pressure is set by capillarity, \(\delta p \sim\gamma/\lambda\), and Darcy flow gives
\(M\sim k_0 c_s/\eta_D\) with \(\nabla\mu=(1/c_s)\nabla p\). Heree, $c_s$ is the solvent concentration in the gel where it is transported by coarsening of the solvent-rich domain.
For slow poroelastic spreading, the exact diffusion solution yields
\(\left.\nabla p\right|_{r=\lambda}\sim \delta p/\ell_p\sim (\gamma/\lambda)/\ell_p\),
so that
\begin{equation}
	\dot{\lambda}\sim - \frac{1}{\Delta c_s}\,\frac{k_0 c_s}{\eta_D}\left.\nabla p\right|_{r=\lambda}
	\sim - \frac{k_0}{\eta_D}\,\frac{c_s}{\Delta c_s}\,\frac{\gamma}{\lambda\,\ell_p(t)},
	\qquad
	\ell_p(t)=\sqrt{D_p t},\ \ D_p\sim \frac{k_0 G}{\eta_{ps}}.
\label{eq:lambda_dot_gel_slow_app}
\end{equation}
For scaling purposes, the dimensionless ratio \(c_s/\Delta c_s\) is  treated as a constant of order unity and we omit its numerical value here.

\subsubsection{Gel (poroelastic; fast poroelastic diffusion)}
Once poroelastic diffusion has spread over distances comparable to the domain size, the
interfacial gradient saturates to \(\left.\nabla p\right|_{r=\lambda}\sim \delta p/\lambda\sim \gamma/\lambda^2\) (see Eq.~\ref{eq:gradp_interface_app}),
and the evolution law becomes
\begin{equation}
	\dot{\lambda}\sim - \,\frac{k_0 }{\eta_D}\left.\nabla p\right|_{r=\lambda}
	\sim - \frac{k_0}{\eta_D}\,\,\frac{\gamma}{\lambda^2}.
\label{eq:lambda_dot_gel_fast_app}
\end{equation}
This has the same \(\lambda^{-2}\) structure as LSW, but with the molecular mobility
\(D\) replaced by the Darcy permeation mobility \(k_0 c_s/\eta_D\), and with the additional
dimensionless factor \(c_s/\Delta c_s\) which we omitted for scaling purposes.

Depending on whether the interfacial pressure gradient has reached its saturated value
\(\left.\nabla p\right|_{r=\lambda} \sim \delta p/\lambda\) or is still building up in time as
\(\left.\nabla p\right|_{r=\lambda} \sim \delta p/\ell_p(t)\),
with $\ell_p=\sqrt{D_p \, t}$, one recovers either the standard LSW coarsening law $t^{1/3}$ or the slow poroelastic coarsening law $t^{1/4}$
discussed in the main text.

\section{Mechanical arrest  of pressure gradients in a gel in its elastic regime} \label{app:spherical_inclusion_elastic_bc}

In this appendix we analyze the role of pressure gradients in a solvent-rich domain embedded in an \textit{elastic} gel. We show that in the elastic regime, force balance at the solvent-rich domain interface with the gel, is established by the balance of the elastic stress and the Laplace pressure.  This results in a solvent pressure gradient which is zero, thus arresting Darcy flow of the solvent and hence, coarsening.

We model the gel as a two-component  medium consisting of an \textit{elastic} polymer network and an incompressible solvent. The total stress is
\begin{equation}
	\boldsymbol{\sigma} = -p\,\mathbf{I} + \boldsymbol{\sigma}^{\text{el}},
    \label{eq:D1}
\end{equation}
where $p$ is the solvent pressure and $\boldsymbol{\sigma}^{\text{el}}$ is the elastic stress of the polymer network.

For a linear isotropic elastic solid,
\begin{equation}
	\boldsymbol{\sigma}^{\text{el}} = K (\nabla \cdot \mathbf{u})\,\mathbf{I} + 2G\,\boldsymbol{\varepsilon}^{\text{dev}},
\end{equation}
where $\mathbf{u}$ is the displacement field and $\boldsymbol{\varepsilon}^{\text{dev}}$ is the deviatoric strain \cite{LandauLifshitz}.
Mechanical equilibrium (force balance) requires:
\begin{equation}
	-\nabla p + \nabla \cdot \boldsymbol{\sigma}^{\text{el}} = 0
\end{equation}

When the elastic stress is divergence-free:
\begin{equation}
	\nabla \cdot \boldsymbol{\sigma}^{\text{el}} = 0
    \label{eq:D4}
\end{equation}
force balance implies
\begin{equation}
	\nabla p = 0
    \label{eq:D5}
\end{equation}
so the pressure is spatially uniform and the same in the solvent-rich domain and in the solvent pores of the gel.  A divergence-free elastic stress corresponds to mechanical equilibrium in the absence of body forces
and to the minimization of the elastic energy of the gel in the absence of body forces. Thus, the pressure in the solvent-filled regions of the gel vanishes in the elastic regime, even though the gel remains swollen with solvent-filled pores. 

We now explicitly solve for the displacement field in spherical symmetry. The radial component of the displacement field is: 
$\mathbf{u} = u(r)\,\hat{\mathbf{r}}$.
The local volume change in the elastic gel is the divergence of the displacement \cite{LandauLifshitz}:
\begin{equation}
	\nabla \cdot \mathbf{u} = \frac{1}{r^2}\frac{d}{dr}(r^2 u).
\end{equation}
The general spherically symmetric solution to the elastic equilibrium,  Eq.~\ref{eq:D4} is
\begin{equation}
	u(r) = \frac{A}{r^2} + Br
\end{equation}
Requiring that the displacement remain finite at large $r$ fixes $B=0$ so
\begin{equation}
	u(r) = \frac{A}{r^2}.
\end{equation}
The amplitude $A$ of the displacement field is fixed by the mechanical boundary condition at the surface of the solvent-rich droplet. At the interface $r = R$, the normal stress in the gel must balance the pressure inside the droplet.

The total radial stress in the gel is
\begin{equation}
	\sigma_{rr} = -p + \sigma^{\text{el}}_{rr}
\end{equation}
For the uniform-pressure solution ($\nabla p = 0$ from Eq.~{eq:D5}), the pressure $p$ of the solvent in the gel is constant and the pressure in the gel is spatially uniform, but differs from the droplet pressure only by the Laplace jump. Because the solvent inside the droplet and in the pores of the gel are the same fluid phase, the only pressure discontinuity is the localized Laplace jump at the curved interface. Outside the interface, the solvent pressure in the gel is spatially uniform and the only pressure discontinuity is the capillary Laplace jump localized at the curved interface.  Thus, stress variation arises entirely from the elastic contribution. The boundary condition at $r = R$ is
\begin{equation}
	\sigma_{rr}(R) = -p_{\text{in}}
\end{equation}
where $p_{\text{in}}$ is the pressure inside the droplet.

For a purely deviatoric displacement field
\begin{equation}
	u(r) = \frac{A}{r^2}
    \label{eq:D11}
\end{equation}
the radial strain is
\begin{equation}
	\varepsilon_{rr} = \frac{du}{dr} = -\frac{2A}{r^3}
\end{equation}
and the tangential strain is
\begin{equation}
	\varepsilon_{\theta\theta} = \varepsilon_{\phi\phi} = \frac{u}{r} = \frac{A}{r^3}
\end{equation}

Since the deformation is traceless,
\begin{equation}
	\varepsilon_{rr} + 2\varepsilon_{\theta\theta} = 0
\end{equation}
the stress is purely deviatoric, and the radial elastic stress is
\begin{equation}
	\sigma^{\text{el}}_{rr} = 2G \varepsilon_{rr} = -\frac{4G A}{r^3}
\end{equation}
Evaluating the stress at the interface $r = R$ gives
\begin{equation}
	\sigma^{\text{el}}_{rr}(R) = -\frac{4G A}{R^3}.
\end{equation}

Force balance requires that the normal stress is continuous at the interface  gives,
\begin{equation}
	-p + \sigma^{\text{el}}_{rr}(R) = -p_{\text{in}}
\end{equation}
where the right hand side is the stress within the drop, \textit{including} the Laplace pressure. 
For the zero pressure-gradient solution, the pressure in the gel is spatially uniform, and differs from the droplet pressure \textit{only} by the Laplace jump at the curved interface. We thus obtain
\begin{equation}
	-\frac{4G A}{R^3} = 2\frac{\gamma}{R}
\end{equation}

If the pressure difference across the interface is set by surface tension $\gamma$,
\begin{equation}
	p_{\text{in}} - p = \frac{2\gamma}{R},
\end{equation}
then
\begin{equation}
	A =- \frac{\gamma R^2}{2G}
    \label{eq:D20}
\end{equation}
the negative value of A corresponds to an inward radial displacement  of the gel near the droplet due to the inward capillary stress at the curved interface. The resulting strain field is purely deviatoric: the inward displacement produces a purely deviatoric strain field: circumferential directions are compressed while the radial strain is tensile, with zero net volume change.
The amplitude $A$ is determined by the balance between interfacial forces (Laplace pressure) and the elastic resistance of the surrounding network.

In the main text, we denote the radial displacement at the boundary of the gel and the solvent-rich domain as $\delta \lambda=u(R)$ where $u(R)$ is given by Eq.~\ref{eq:D11} evaluated at $r=R$ and with $A$ determined by force balance, Eq.~\ref{eq:D20}.  Thus $\delta \lambda=-\gamma/(2 G)$ which is of order of the mesh-size and a small displacement relative to the micron-sized solvent domain, $\lambda$.

For this solution,
\begin{equation}
	\nabla \cdot \mathbf{u} = 0,
\end{equation}
so the deformation is purely deviatoric  (no volumetric strain) and satisfies
\begin{equation}
	\nabla \cdot \boldsymbol{\sigma}^{\text{el}} = 0,
\end{equation}
By Eq.~\ref{eq:D1}, this corresponds to a state with a zero pressure gradient (uniform pressure). There is thus no driving force for Darcy flow, which arrests coarsening at the time $\tau_{el}$ at which the gel enters its elastic regime. Arrest of coarsening (due to $\nabla p=0$) is due to the force balance at the drop being established by the elastic stress and capillary pressure.  This is only possible in the elastic regime; in the viscoelastic regime of the gel the elastic stress is zero and force balance at the drop interface gives rise to the pressure gradient driving Darcy flow and hence coarsening.

 \newpage

\section{Experimental Scaling of Arrest Length} \label{app:lambdastar}

\begin{figure}[h]
    \centering
    \includegraphics[width=1\linewidth]{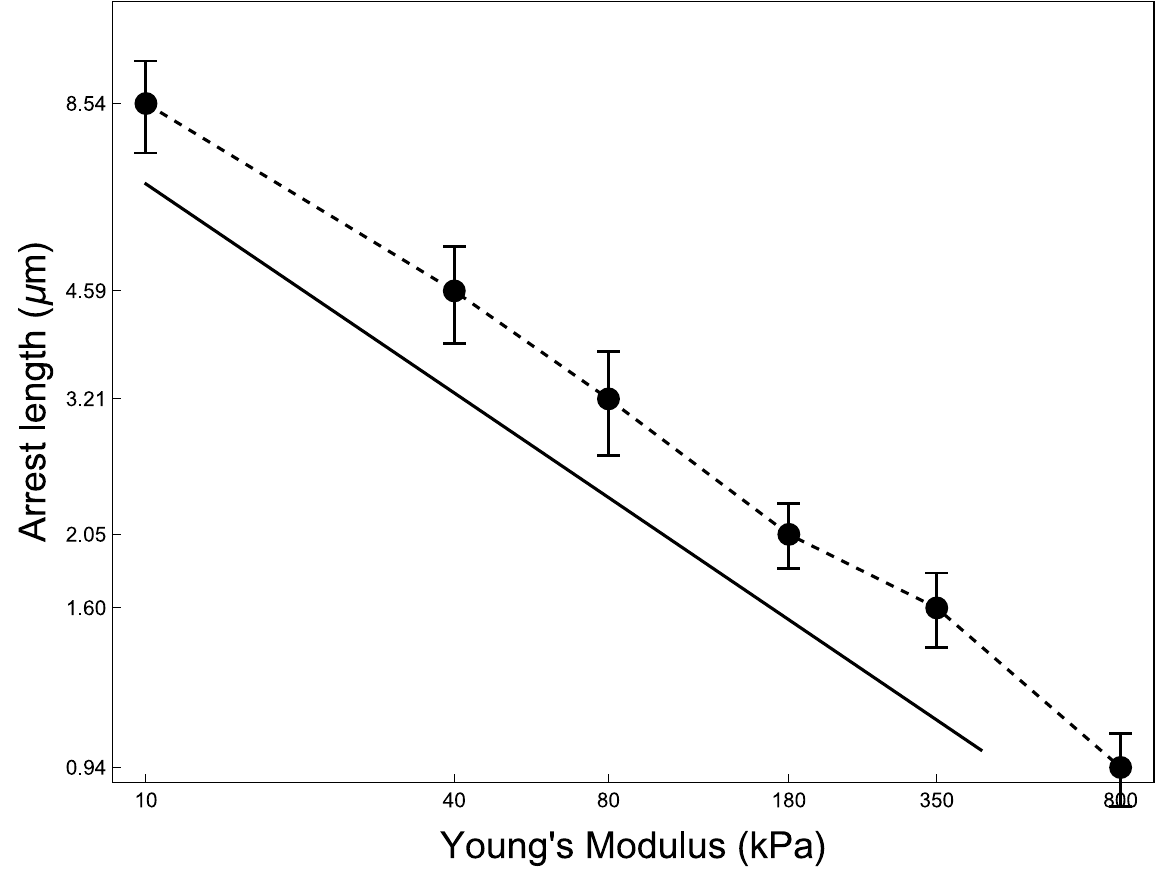}
    \caption{Experimental measurements of domain size ($\lambda^*$ in  the notation of Ref. 1, arrest length in our theory)   as a function of the Young’s modulus on a log-log plot.  In our scaling theory, in the slow poroelastic regime, the domain size at arrest is predicted to vary with the gel modulus $G$ as $1/\sqrt{G}$ which is indicated by the solid line in the figure, in agreement with experiment. The dashed line is a guide to  the eye.  [Redrawn with data from Figure 3D taken from Ref. 1.]}
    \label{fig:Dufresne1}
\end{figure}

\newpage

\bibliographystyle{apsrev4-2}
\bibliography{ScalingRefs}

\end{document}